# Observation of decoupling of electrons from phonon bath close to a correlation driven metal-insulator transition

Sudipta Chatterjee[1,#] , Ravindra Singh Bisht[1,#], V.R.Reddy[2], A.K. Raychaudhuri[3,*]

[1] Department of Condensed Matter Physics and Materials Science,
S.N. Bose National Centre for Basic Sciences, JD Block, Sector III, Salt Lake,
Kolkata 700106, INDIA

[2] UGC-DAE-Consortium for Scientific Research, University Campus, Khandwa Road,
Indore 452017, INDIA

[3] CSIR Centre for Glass and Ceramic Research Institute, 196 Raja S.C. Mullick Road,
Kolkata 700032, INDIA

*corresponding author, Email: arupraychaudhuri4217@gmail.com

# Both authors made equal contributions

**Abstract**

We observed that close to a Mott transition, over a small temperature range, the predominance of slow relaxations leads to decoupling of electrons from the thermal bath. This has been established by observation of large deviation of the thermal noise in the films of Mott system $NdNiO_3$ from the canonical Johnson-Nyquist value of $4k_BTR$ close to the transition. It is suggested that such a large noise arise from small pockets of nanometric metallic phases (estimated size ≈ 15-20nm) within the insulating phase with the Coulomb charging energy as the control parameter.

Metal-Insulator Transition (MIT), where a metal with delocalized electrons makes a transition to an insulating state with strongly localized electrons, is one of the most fascinating areas of research in modern condensed matter physics. Despite seminal contributions that enriched this field, there are fundamental unresolved issues that need attention. The most attractive feature of MIT is that it is a general phenomenon observed in a number of systems and despite the differences, there are certain attributes that have ubiquitous presence in the physics of MIT.

In recent years, the exciting issue of slowing down of charge relaxation close to MIT have been investigated . Critical slowing down has been observed through Noise measurements[1,2] as well as in NMR relaxation time measurements[3,4,] in polymeric /organic conductors as well as in oxides like in $V_2O_3$[2] undergoing Mott transition. Slow kinetics close to the Mott transition manifests as large correlated flicker noise (Spectral Power density $S_V(f) \propto \frac{1}{f^\alpha}$) with non-Gaussian characteristics have been reported in oxides like $VO_2$ [5] and $NdNiO_3$[6,7,8,9] that show Mott Transition. . In this paper, we report a fundamental aspect of the temperature driven Mott transition that has not been reported before which has significant thermodynamic consequences. We observe a new phenomena that thermal noise with very large spectral power density ($S_{th}$) , appears at a temperature range close to the MIT temperature ($T_{MI}$) , concomitant with the appearance of large correlated flicker noise. The observed $S_{th}$ is significantly higher than the Johonson-Nyquist value[10,11] showing decoupling of the electrons from the lattice thermal bath.

The experiment was done in films of $NdNiO_3$ grown on crystalline $SrTiO_3$ (STO) substrates of different crystallographic orientations. $NdNiO_3$ (NNO) undergoes temperature-driven Mott type MIT that has attracted considerable attention in recent years[12,13]. Though investigated on rare-earth nickelate $NdNiO_3$, the reported phenomenon as described is of general validity and is expected to be observed in any Mott transition.

The field of MIT developed along two directions. One direction is the disorder driven Anderson transition[14,15,16] where the density of states (DOS) at the Fermi level ($N(E_F)$) remains finite ($N(E_F) \neq 0$) although with localized electronic states around $E_F$. The other direction is the correlation driven Mott transition[17] where the DOS at $E_F$ splits into two bands with $N(E_F) \rightarrow 0$ at the transition. In recent years there is a convergence of the two broad classes of MIT which is referred as the Mott-Anderson transition where there is a presence of both disorder and correlation [18,19,20]. It has been shown that the DOS at $E \approx E_F$ gets significantly modified by presence of disorder in a Mott-Transition[21]. The present report is placed in this contextual framework.

The thermodynamics of MIT has important physical consequences. While Anderson Transition is a continuous transition[16], the Mott transition is generally thought of as 1st order transition which can be broadened by disorder[1-4,18]. In MIT, like in any other phase transition, one would expect critical slowing down close to critical region. For Anderson transition, such slowing down has been observed at $T \rightarrow 0$ through resistance/conductance noise measurements in 2-dimensional systems like MOSFET Si inversion layers[22,23,24] and in 3-dimensional system Si (P, B)[25]. Slow relaxation in the Mott transition region [1-4] are finite temperature analogues of the phenomena observed near $T \rightarrow 0$ in Anderson transition.

Thermal noise[10,11] is a consequence of the Fluctuation Dissipation Theorem (FDT)[26] and shows up as a voltage fluctuation (without an applied bias) across a dissipative circuit element like a resistor $R$ kept at a bath temperature $T$. The mean square voltage fluctuation $\langle(\delta V)^2\rangle$ measured over a bandwidth $\Delta f$ is given by the relation[10,11] :

$$\langle(\delta V)^2\rangle = (4k_B TR)\Delta f \tag{1}$$

For Eqn. 1 it is assumed that the electron system, with temperature $T_e$ is in equilibrium with the phonon thermal bath at temperature $T$ so that $T_e = T$. The fluctuation gives a fundamental measurement of the bath temperature $T$ and forms the basis of noise thermometry in metrology[27]. Equilibrium Johnson- Nyquist thermal noise has a frequency-independent spectral power density (SPD) given as[10,11]:

$$S_{th} = 4k_B TR \tag{2}$$

Eqn. 2 is applied to measure the temperature $T_e$ in a hot electron system where measured thermal noise $S_{th} > 4k_B RT$ and $T_e > T$ [28]. The thermal noise co-exists with the flicker noise that has SPD $S_V(f) \propto \frac{1}{f^\alpha}$ often referred to as "$\frac{1}{f}$" noise.

The NNO films of thickness 15nm and root mean square (RMS) roughness of 0.3nm were grown on crystalline STO substrate with different crystallographic orientations by Pulsed Laser Deposition (PLD) using a KrF (λ = 248nm) laser. Some of the details of growth and characterization of the films are given in previous publication from the group[7]. The films are strain relaxed although with residual strain (see Table I) due to finite mismatch of lattice constants of the substrate and the film, as established by the reciprocal space mapping (RSM). The strain relaxation creates coherent grains of average size $\geq 35 - 40nm$, as seen from Atomic Force Microscopy. Strain relaxation and formation of misfit dislocations make the strain inhomogeneous that modulates the nature the electronic phase separation (EPS) leading to co-existing phases. (The RSM data in Supplementary Information. Figures S1-S6 Table SI .)

The resistivity ($\rho$) was measured down to 3K. The noise measurements (for 80K < $T < 300K$) were performed using a 5-probe a.c. excitation[29] technique using temperature stabilization of ±5 mK. This method[29] allows simultaneous measurements of the flicker noise (SPD $S_V(f) \propto \frac{1}{f^{\alpha}}$) as well as the frequency-independent thermal noise (SPD $S_{th}$). (Details in Supplementary Information and elsewhere[30,31]). A number of extraneous noise sources can add to the observed $S_{th}$ and make it deviate from the canonical value $4k_BTR$ (Eqn. 2) . Proper elimination of extraneous factors allowed us to reach a noise floor on the ratio $\zeta \equiv \frac{S_{th}}{4k_BTR} \leq 1.5$.

In Figure 1 we show an example of the frequency-independent thermal noise $S_{th}$ along with the flicker noise with SPD $S_V(f) \propto \frac{1}{f^{\alpha}}$. taken on a film of NNO grown on STO with (111) orientation (NNO/STO (111)). $\rho$ $vs$ $T$ data for heating and cooling cycle are in the inset.

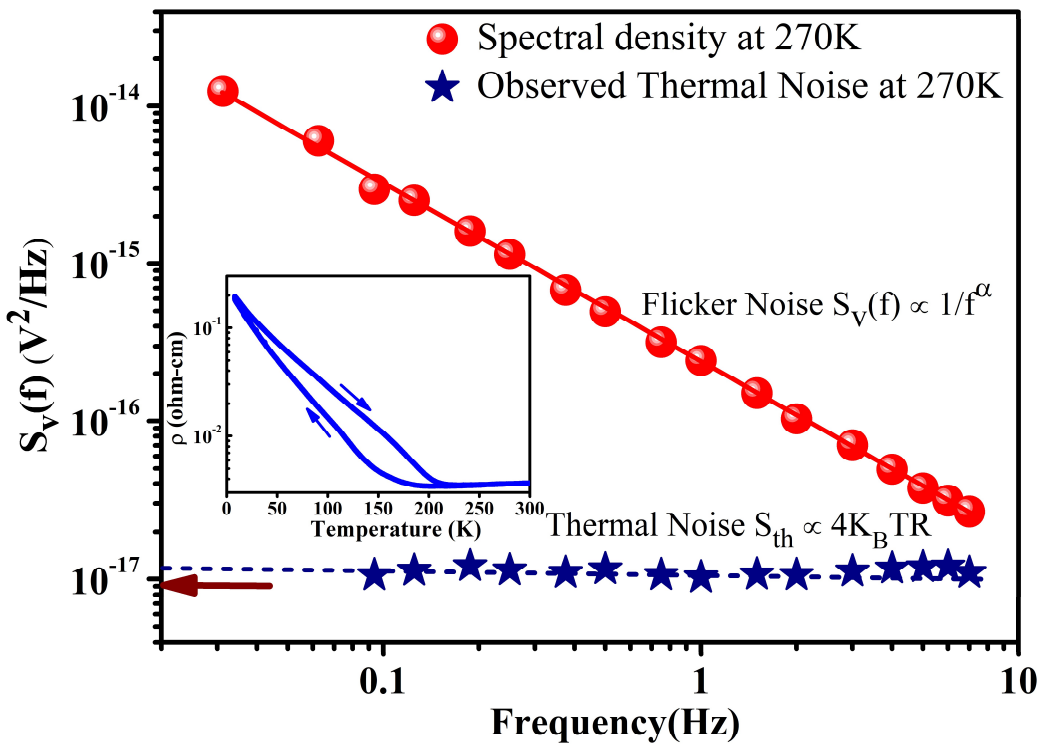

**Figure 1**: Example of measured SPD ($S_V$ (f)) for the flicker noise varying as $\frac{1}{f^{\alpha}}$ and measured thermal noise $S_{th}$ (white spectrum) in a NNO/STO (111) film . The Nyquist- Johnson value $4k_B$TR at $T = 270K$ marked by an arrow. Inset : Resistivity of the film. Resistivity data on other films in Supplementary Figure S7.

The MIT temperatures ( $T_{MI}$) , determined from the change in the sign of the derivative $\frac{d\rho}{dT}$ ( see supplementary data Figure S8-S10), have been summarized in Table I for all the NNO films.. In the transition range, EPS leads to co-existence of the insulating and metallic phases as observed by spatially resolved techniques [7,32].

The dependence of the fractions of co-existing phases evaluated from the $\rho - T$ curves using effective medium theory[33] are shown in supplementary Figures S11-S13.The insulating fraction $f_i$ at $T = T_{MI}$ is $\approx .09 - .11$ for all the films. The resistivity curves show hysteresis for heating and cooling cycles. The hysteresis happens in a Mott transition due to its underlying first order nature and can be tuned by extent of co-existing phase fractions[34,35] as shown in systems like $PrNiO_3$ and $V_2O_3$. The hysteresis , we show below also persists in the noise data that show distinct dependence on heating and cooling cycles.

Table I: Compilation of relevant experimental data

| **Sample** | $\epsilon_\perp(\%)^{\#}$ | $\epsilon_\parallel(\%)^{\$}$ | **$T_{MI}$ (K)** | | **$T^*$ (K)** | | **$T^*/T_{MI}$** | | $\zeta_M$ | |
|---|---|---|---|---|---|---|---|---|---|---|
| | | | **H** | **C** | **H** | **C** | **H** | **C** | **H** | **C** |
| NNO/STO(100) | -0.081 | 0.094 | 177 | 160 | 160 | 142 | 0.90 | 0.89 | 10 | 4 |
| NNO/STO(110) | 0.015 | -0.017 | 206 | 189 | 200 | 182 | 0.97 | 0.96 | 11 | 7 |
| NNO/STO(111) | 0.359 | -0.419 | 211 | 187 | 228 | 202 | 1.08 | 1.08 | 12 | 10 |

# :, Out- of -plane strain (c-axis) $\epsilon_\perp = \frac{c_{film-}c^R_{film}}{c_{substrate}}$. $:In-plane strain (a-axis) $\epsilon_\parallel = \frac{a_{film-}a^R_{film}}{a_{substrate}}$ as determined from X-ray data. Subscripts film and substrate refer to the film and the substrate respectively. The super-script R refers to the fully relaxed film. H = Heating cycle, C = Cooling Cycle, $T_{MI}$ = MI transition temperature, $T^*$= Temperature at which thermal noise ($S_{th}$) shows a peak. $\zeta_M$ is the peak value of the ratio $\zeta \equiv \frac{S_{th}(\mathrm{T})}{4k_BTR}$ observed at $T^*$.

In Figure 2 we shows $\zeta(T) \equiv \frac{S_{th}(\mathrm{T})}{4k_BTR}$ which is the measured thermal noise $S_{th}$(T), scaled by canonical Johnson Nyquist (JN) value $4k_BTR$ for the NNO films (kept at temperature $T$) grown on different STO substrates during heating cycle . (Cooling cycle data in Figure S14). This is the most important result which shows the decoupling of the electrons from the lattice thermal bath. For a system in thermal equilibrium theoretically $\zeta \approx 1$. (The extraneous noise limit is $\zeta \leq 1.5$ as stated before.) We find that in a narrow temperature range close $T_{MI}$ $\zeta \gg 1$. The measured $\zeta$ reaches a peak (referred as $\zeta_M$ ) at some temperature that we refer to as $T^*(\neq T_{MI})$. $S_{th}$ reaches a maximum at $T^*$ and stays high over a range temperature around $T^*$. For the heating cycle for all the films $\zeta_M \geq 10$ and during the cooling cycle $\zeta_M \geq 4$. $T^*$ is also larger in the heating cycle compared to that in cooling cycle.

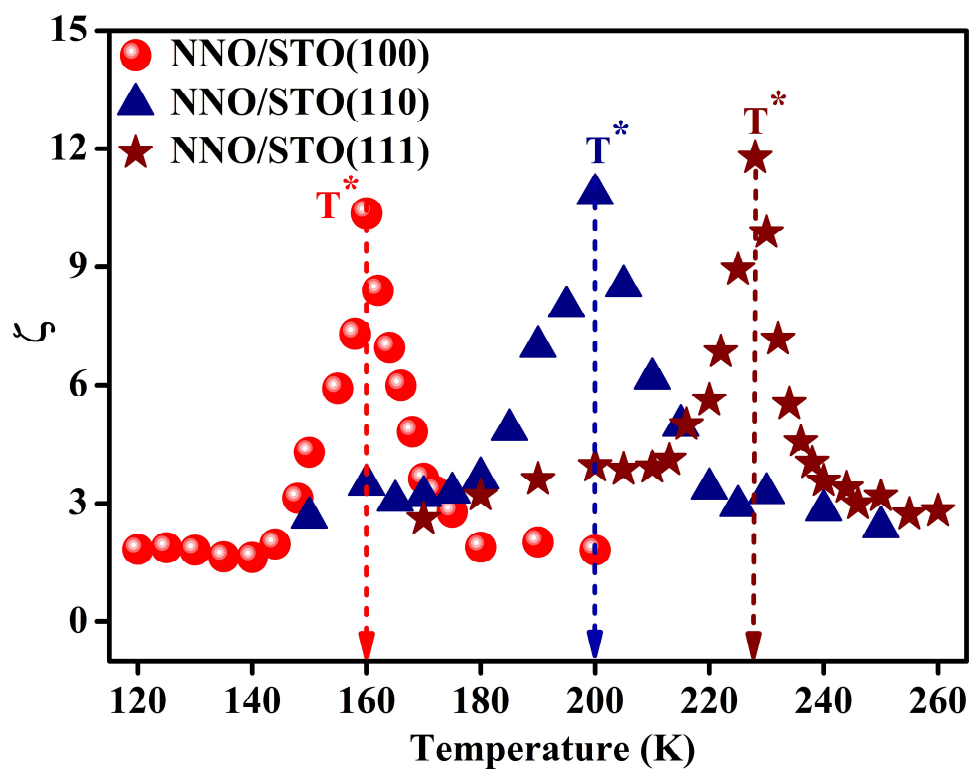

**Figure 2.** The temperature dependence of $\zeta (\equiv \frac{S_{th}(T)}{4k_B TR})$ for NNO films grown on STO substrates with different orientations. Data are shown for the heating cycle. The arrows mark $T^*$. Cooling cycle data in Figure S14

The appearance of large $S_{th}$ near MIT is accompanied by a concomitant appearance of large low frequency flicker noise arising from slowing down of the fluctuation kinetics. This is shown in Figure 3 that shows the temperature variation of the exponent $\alpha$ for the film NNO /STO (111) for both heating and cooling cycles. At $T \sim T^*$ , $\alpha$ deviates significantly from the expected value of $\alpha \approx 1.0 \pm 0.1$ and like $\zeta$ reaches a peak at $T \approx T^*$. The relative variance of the resistance fluctuation ($\frac{\langle \Delta R^2 \rangle}{R^2}$) (Figure 3 inset ) also becomes large reaching a peak at $T \approx T^*$. $\frac{\langle \Delta R^2 \rangle}{R^2} \equiv \int_{f_{min}}^{f_{max}} df \left(\frac{S_V(f)}{V^2}\right)$ evaluated over the bandwidth of the measurement $f_{max}, f_{min}$ . It is noted that the peaks in $\alpha$ and $\frac{\langle \Delta R^2 \rangle}{R^2}$ do not occur at the MIT temperature $T_{MI}$ but at $T^*$. Data for other films in Figures S15 and S16.

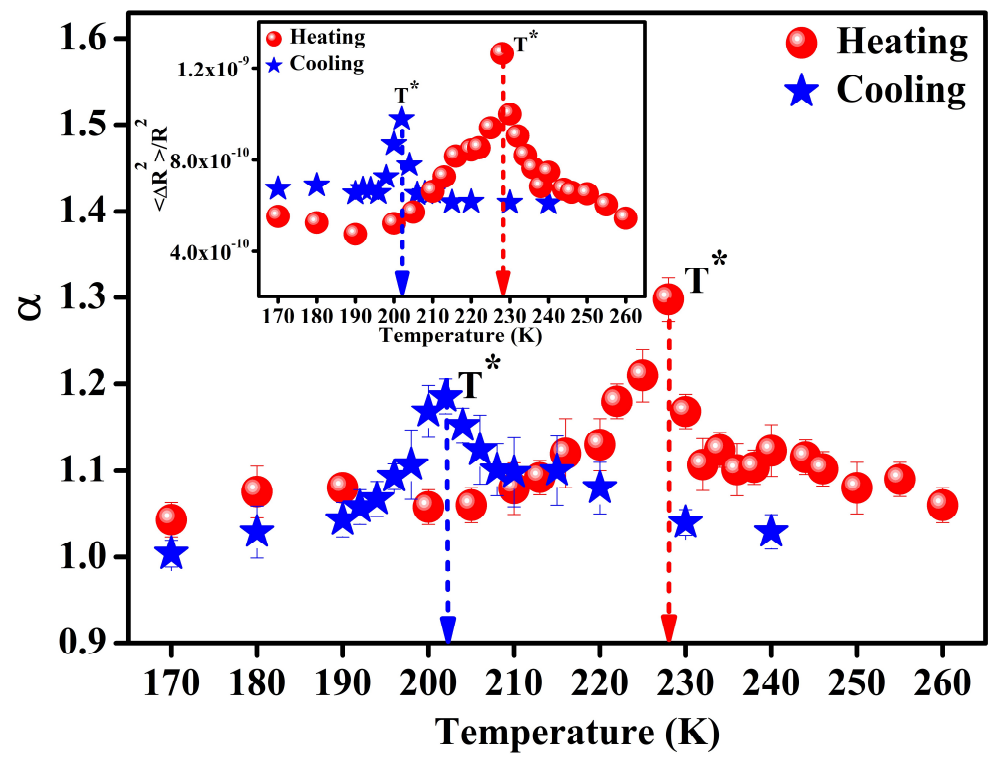

**Figure 3.** Temperature variation of the exponent $\alpha$ for the Flicker noise, $S_V \sim \frac{1}{f^{\alpha}}$, for NNO/STO (111) film. Inset: Relative variance $\frac{\langle \Delta R^2 \rangle}{R^2}$. The arrow marks $T^*$.

The appearance of large $S_{th}$ is correlated with correlation time of fluctuation ($\tau$) as obtained from the autocorrelation function $\left(C(t)\right)$ of the voltage fluctuations (that gives the flicker noise). $C(t)$ is obtained from the times series of fluctuation $\delta v(t)$ by the relation $C(t) \equiv < \delta v(t') \times \delta v(t + t') >_{t'}$, where $<\cdots\cdots>_{t'}$ represents the time average. $C(t)$ shows an approximate exponential dependence for small time and a long time tail. We obtain the correlation time $\tau$ approximating it as the time when $C(t) = \frac{1}{e} C(0)$ .The correlation time ($\tau$) reaches a maxima at $T = T^*$ for the film NNO/STO(111) as shown in Figure 4 . An example of $C(t)$ $at$ $T = T^*$ is shown in the inset at top left for both cycles as examples. (Data for other films in Figure S17 and 18).

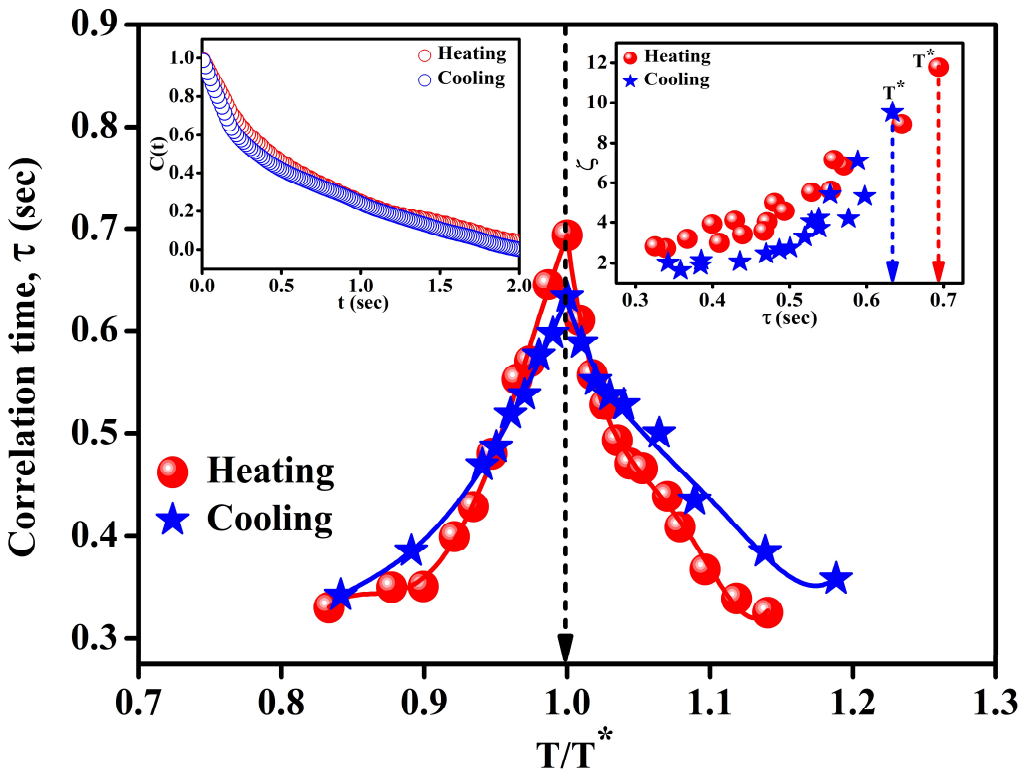

**Figure 4**. The correlation time $\tau$ as a function of scaled temperature $\frac{T}{T^*}$ for film NNO/STO(111) for heating and cooling cycles. The inset at left top corner shows the auto-correlation function $C(t)$ vs. $t$ for NNO/STO(111) at $T = T^*$. Inset at right upper corner shows a plot of $\zeta$ vs $\tau$ for the film NNO/STO(111) .

A plot of $\zeta$ as a function of $\tau$ shown in the inset at right upper corner of Figure 4 shows a direct correlation of the two thus establishing that the decoupling of the electrons with the thermal bath occurs when there is slowing down of the time scales of fluctuation.

At and around $T^*$, the fluctuation also becomes strongly correlated and develops a component of non-Gaussianity as measured by the normalized second spectrum of the fluctuation ($\Gamma^2$). Data and details are given in Supplementary Information Figures S19-S21.

As stated before, previous reports of noise spectroscopy in polymeric conductors[1,3] as well as in transition metal oxides[2,5-9] have reported emergence of low-frequency correlated fluctuations near the MIT as observed from exponent $\alpha$, variance $\frac{\langle \Delta R^2 \rangle}{R^2}$ and second spectra $\Gamma^2$. However, no report exists that observes a large deviation of the thermal noise from the JN value signifying decoupling of electron bath from the phonon bath.

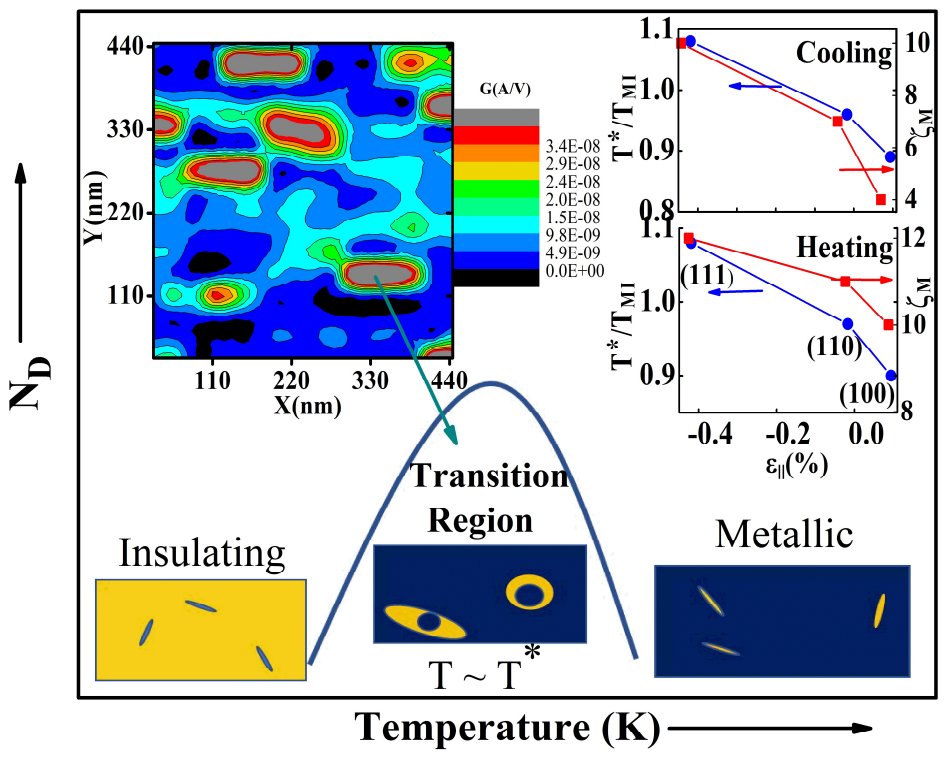

**Figure 5.** Schematic of the temperature variation of $N_D$ in the transition region. The insets show schematic of co-existing phases. Dependence of $\frac{T^*}{T_{MI}}$ and $\zeta_M$ and on in-plane strain $\epsilon_{\|}(\%)$ for films of NNO/STO films with different crystallographic orientations shown in upper right inset . The LCMAP obtained from STM is shown in upper left inset .

*Discussions*

MIT involves a change in the carrier density $(n)$ at $T\sim T_{MI}$ . The large difference in $n$ in the coexisting phases at $T\sim T_{MI}$ causes inhomogeneous current flow leading to large correlated noise. In an EPS system, as in a percolating network, it is established that the flicker noise will be high[36,37] . In this report, it is proposed that the origin of the observation large thermal noise can also be traced to EPS. This issue has not been investigated before.

The thermal noise will be high if there is a sparse phase of nanometric isolated metallic islands (nano-puddles) in the temperature range $T\sim T^*$ where the insulating phase is the minority phase. The physical scenario proposed (Figure 5) is that of a backbone of conducting network (the majority phase) coupled weakly (by tunneling) with the nano-puddles through the intervening minority insulating phase. Such phases have been proposed in the context of Griffiths phase[38] where one observes slow electronic relaxation[4]. The nano-puddles, with small enough average diameter $\langle d \rangle$, would show

Coulomb charging[39] in the temperature region around $T^*$ , if the charging energy $E_C \geq k_B T^*$ The Coulomb charged weakly coupled nano-puddles will have slow relaxation dynamics and would induce fluctuations in the conduction path formed by the majority phase, which would act as source of both flicker as well as excess thermal noise. · Such a scenario has been proposed for understanding noise (and transport) in systems with co-existing majority metallic and weakly coupled dispersed nanoparticle phases [40,41]. The nonequilibrium electron distribution[42] in the Coulomb blockaded nano-puddles acts as a source of heating[43,44] and the polarization fluctuation around such charged nano-puddles lead to large thermal noise[45].

The number density ($N_d$ ) as well as $\langle d \rangle$, of such nano-puddles will depend on $T$ as shown schematically in Figure 5. The maximum noise will appear in the temperature range where $N_d$ is expected to show a broad maxima, which will happen for reasons stated below. On cooling below $T_{MI}$, the metallic phase shrinks and insulating fraction grows to become the majority phase subsuming the nano-puddles. At $T > T_{MI}$ heating leads to further growth of the majority metallic phase which subsumes the remaining insulating phase separating the nano-puddles from the main phase. Thus there is a temperature window close to $T_{MI}$ (and $T^*$) where isolated nano-puddles embedded in minority insulating phase exist and can have high enough $N_d$ contributing to excess noise.

For $T \sim T^*$ , where the noise is large, $\langle d \rangle$ of the Coulomb charged nano-puddles can be estimated from $E_c \equiv \frac{1}{2}\left(\frac{e^2}{C_d}\right) \geq k_B T^*$. ($e$ = elementary charge , $C_d = 2\pi\varepsilon_0\varepsilon_d\langle d \rangle$ = capacitance of metal nano-puddle of diameter $\langle d \rangle$ embedded in a medium with dielectric constant $\varepsilon_d$ and $\varepsilon_0$ =free space permittivity)[47] . An estimate of $\langle d \rangle \leq 15 - 20$ nm was obtained using $\varepsilon_d \approx 5$ , the d.c dielectric constant [48] of $NdNiO_3$. using the observed $T^*$. $\langle d \rangle$ ≤20nm for NNO/STO(100) and decreases to ≤ 15 nm for NNO/STO(111) that has larger $\zeta$ and higher $T^{*}$. Such nano-puddles have been observed in spatially resolved scanning tunneling spectroscopy (STS) data (see Figure 5 left upper inset) that shows the local tunneling conductance map (LCMAP) and also in past studies[7,32] . Note on LCMAP in Supplementary. (We note that the estimate of $\langle d \rangle$ has been made assuming spherical shape for the puddles which in reality may have ellipsoidal shape.)

The size scales associated with the EPS and hence nano-puddles are modulated by the residual strain field inhomogeneity created by misfit dislocations[48,49,50,51] . The observed dependence of $T^{*}$ on residual strain $\epsilon_{\|}(\%)$ (Figure 5 upper right inset) will thus arise from the dependence of $\langle d \rangle$ on such parameters,

*Concluding remarks*- We report the first observation of the decoupling of the electron system from the temperature bath in films of $NdNiO_3$ close to its Mott type MI transition. It has been observed, close to the transition temperature, through measurement of the thermal noise which shows a large deviation from the canonical Jonson-Nyquist value of $4k_BTR$ . The large rise in the thermal noise occurs due to the predominance of slow relaxations whose presence has been confirmed by a large rise of the correlation time of fluctuations and also concomitant large enhancements in the mean square resistance fluctuation of the flicker noise as well as of the exponent $\alpha$. A scenario has been proposed where it has been suggested that the large noise arises from rare nanometric small pockets of metallic phases (nano-puddles) which are coupled to the majority metallic phase through a layer of minority insulating phase. STS based imaging shows existence of such regions.

**Acknowledgment:**

The work has been supported by a sponsored project from the Science and Engineering Research Board (SERB), Government of India (ref: EMR/2016/002855/ PHY). AKR wants to thank SERB for Distinguished Fellowship (ref: SB/DF/008/2019). SC acknowledges Satyaki Kundu for his helpful suggestions.

# Supplementary Information

# Decoupling of electrons from phonon bath close to a correlation driven metal-insulator transition

Sudipta Chatterjee[1], Ravindra Singh Bisht[1], V.R.Reddy[2], A.K. Raychaudhuri[3]

[1] Department of Condensed Matter Physics and Materials Science,
S.N. Bose National Centre for Basic Sciences, JD Block, Sector-III, Salt Lake,
Kolkata 700106, INDIA
[2] *UGC-DAE-Consortium for Scientific Research*, University Campus, Khandwa Road,
Indore 452017, INDIA
[3] CSIR Centre for Glass and Ceramic Research Institute, 196 Raja S.C. Mullick Road,
Kolkata 700032, INDIA
**- corresponding author, Email: arupraychaudhuri4217@gmail.com*

## A. Reciprocal Space Mapping (RSM) data and analysis:

The x-ray reciprocal space mapping (RSM) measurements are carried out using Bruker D8-Discover system equipped with Cu $K_{\alpha}$ radiation, Eulerian cradle, Goebel mirror and LynxEye detector. The obtained data is analysed with LEPTOS software considering pseudo-cubic notation for NNO layer (0.3807 nm). Since there is no information about the elastic constants of NNO, the Poisson ratio of 0.3 is considered as adopted from published data (Adam J. Hauser, Evgeny Mikheev, Nelson E. Moreno, Jinwoo Hwang, Jack Y. Zhang, *and* Susanne Stemmer *Correlation between stoichiometry, strain, and metal-insulator transitions of $NdNiO_3$ films* Appl. Phys. Lett. 106, 092104 (2015); https://doi.org/10.1063/1.4914002). The RSM measurements are carried out along both symmetric (002) and asymmetric (103) reflections. The data measured along (103) reflection is shown in Figure-S13, S14, S15. The constructed relaxation triangle (one vertex of the triangle is the substrate peak, the other two vertexes are layer peaks for pseudo-morphic and fully relaxed states) is shown in the figures. The data suggest the fully relaxed states, also reflected in the R (Relax) value. Value more than one could be due to the consideration of approximate elastic constants of the NNO layer. The formulas used for calculating the parameters shown in the table are the following.

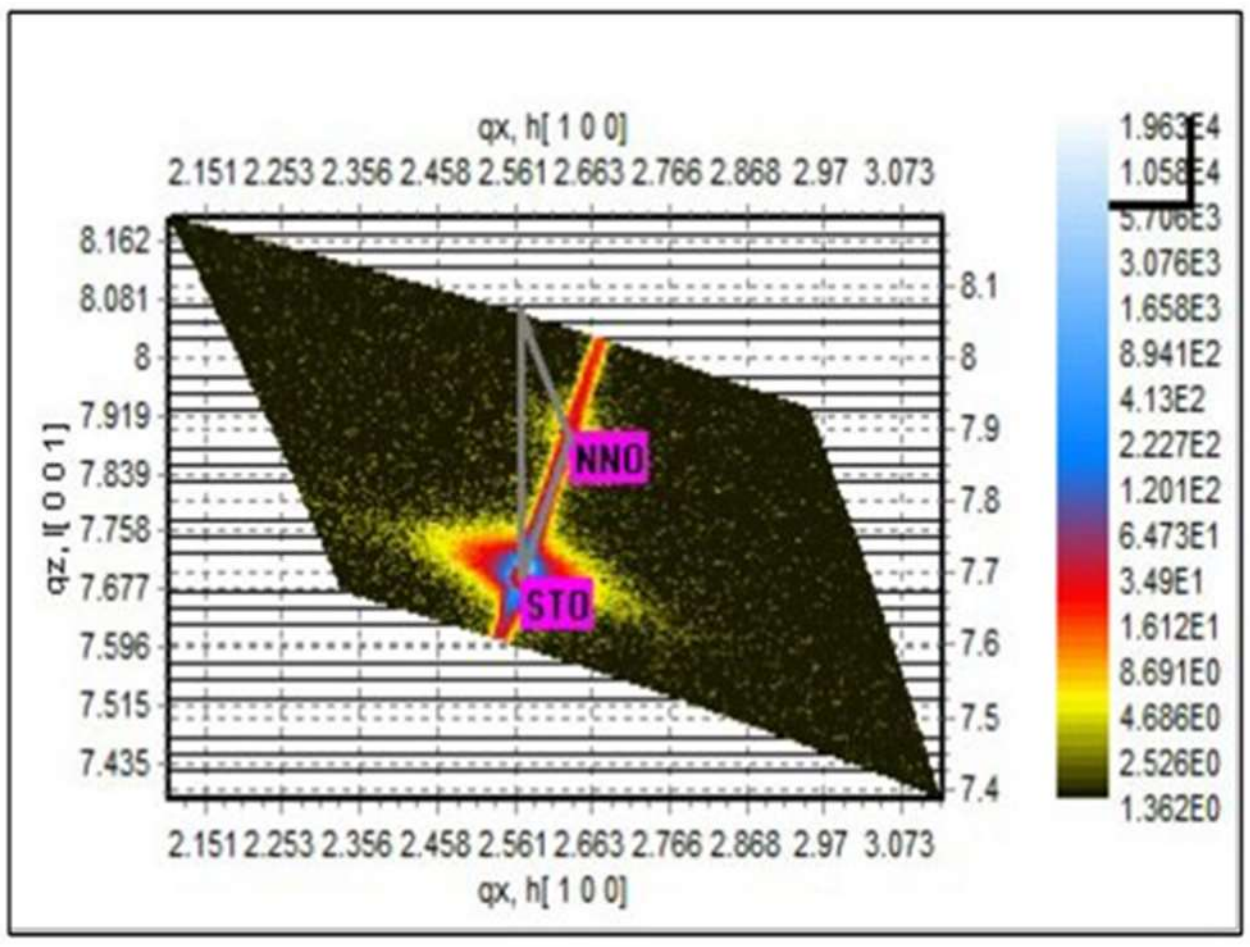

**FIG S1.** RSM data for NNO/STO(100) film measured along (103) reflection. q vectors are in units of $nm^{-1}$

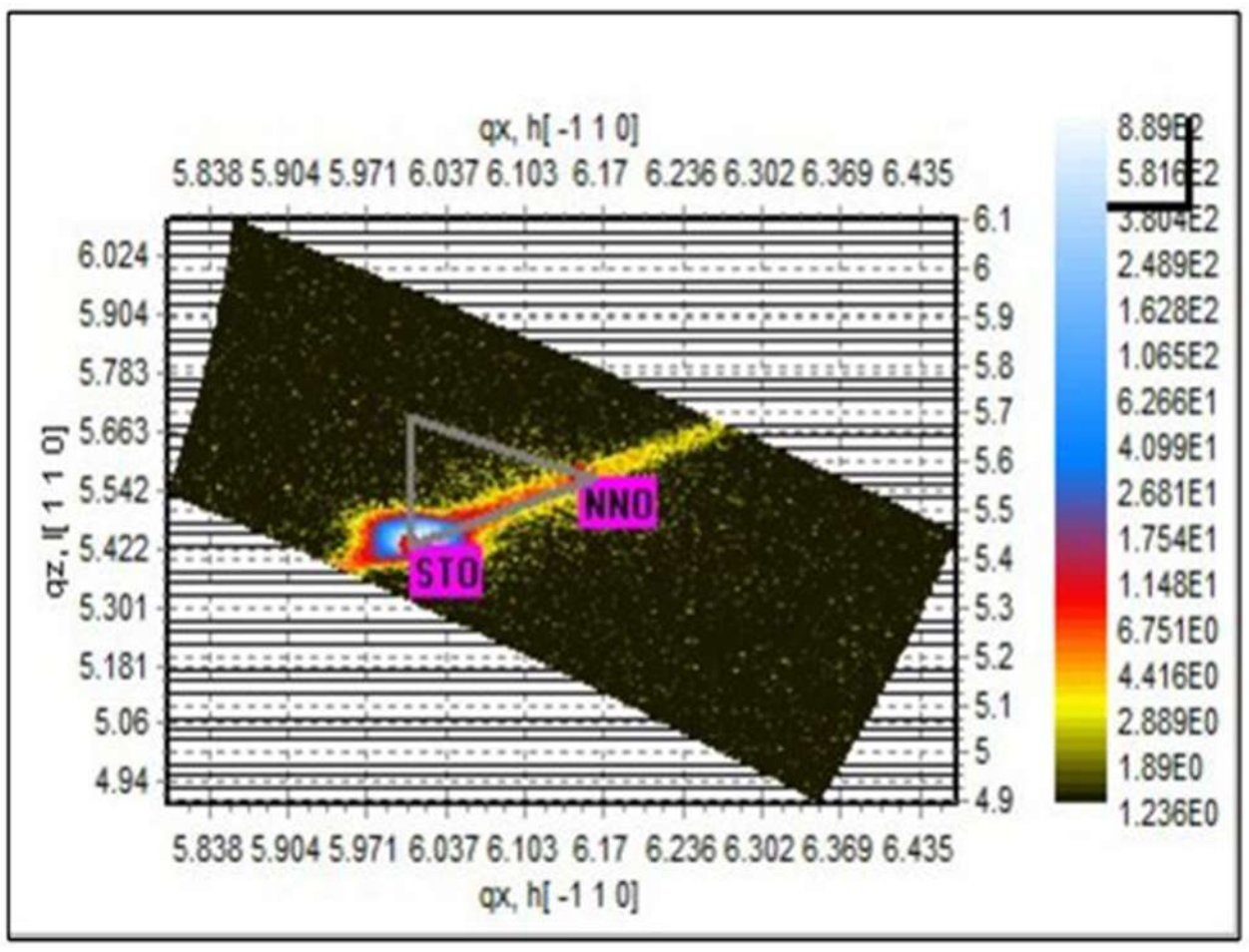

**FIG S2.** RSM data for NNO/STO(110) film measured along (103) reflection. q vectors are in units of $nm^{-1}$

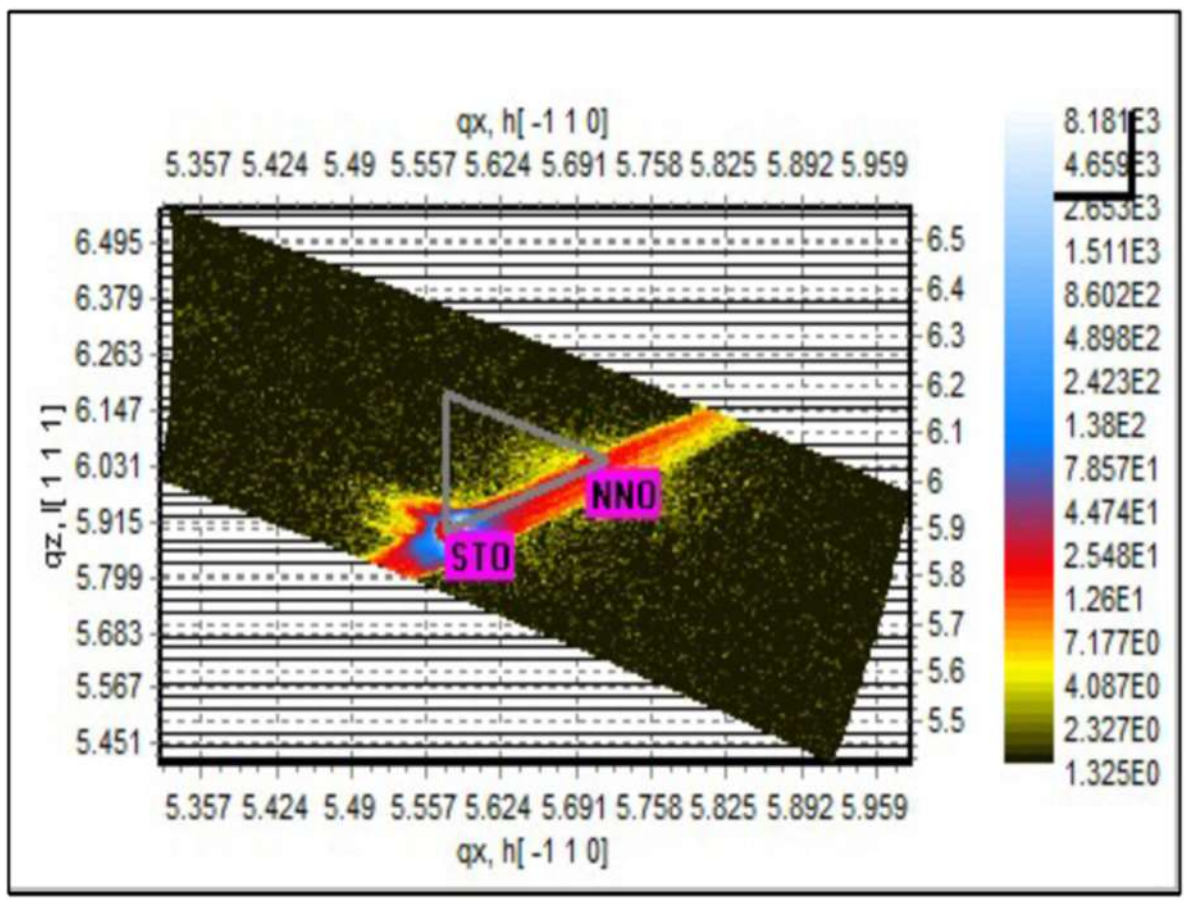

NNO_STO_111_RSM_103 (p1)

**Measured:** 07/20/2019 18:49:48

**FIG S3.** RSM data for NNO/STO(111) film measured along (103) reflection. q vectors are in units of $nm^{-1}$

## Table SI

Important parameters obtained from RSM data

| Sample | $a_{Film}$ (nm) | $c_{Film}$ (nm) | Relax (R) | Δa/a (%) | Δc/c (%) | In-plane Strain % ($\epsilon_{\parallel}$) | Out of plane Strain % ($\epsilon_{\perp}$) |
|---|---|---|---|---|---|---|---|
| NNO/STO(100) | 0.38107 | 0.38038 | 0.96 | -2.415 | -2.592 | 0.094 | -0.081 |
| NNO/STO(110) | 0.38063 | 0.38076 | 1.01 | -2.527 | -2.494 | -0.017 | 0.015 |
| NNO/STO(111) | 0.37906 | 0.38210 | 1.16 | -2.929 | -2.151 | -0.419 | 0.359 |

STO (cubic) : $a,c_{Sub} = 0.3905$ nm

NNO (pseudo-cubic) : $a,c^{R}_{Film} = 0.3807$ nm

$\Delta a/a = (a_{Film}-a_{Sub})/a_{Sub}$

$\Delta c/c = (c_{Film}-c_{Sub})/c_{Sub}$

Lateral strain $= (a_{Film} - a^{R}_{Film})/a_{Sub}$

Vertical strain $= (c_{Film} - c^{R}_{Film})/c_{Sub}$

Relax $= (a_{Film} - a_{Sub})/(a^{R}_{Film} - a_{Sub})$

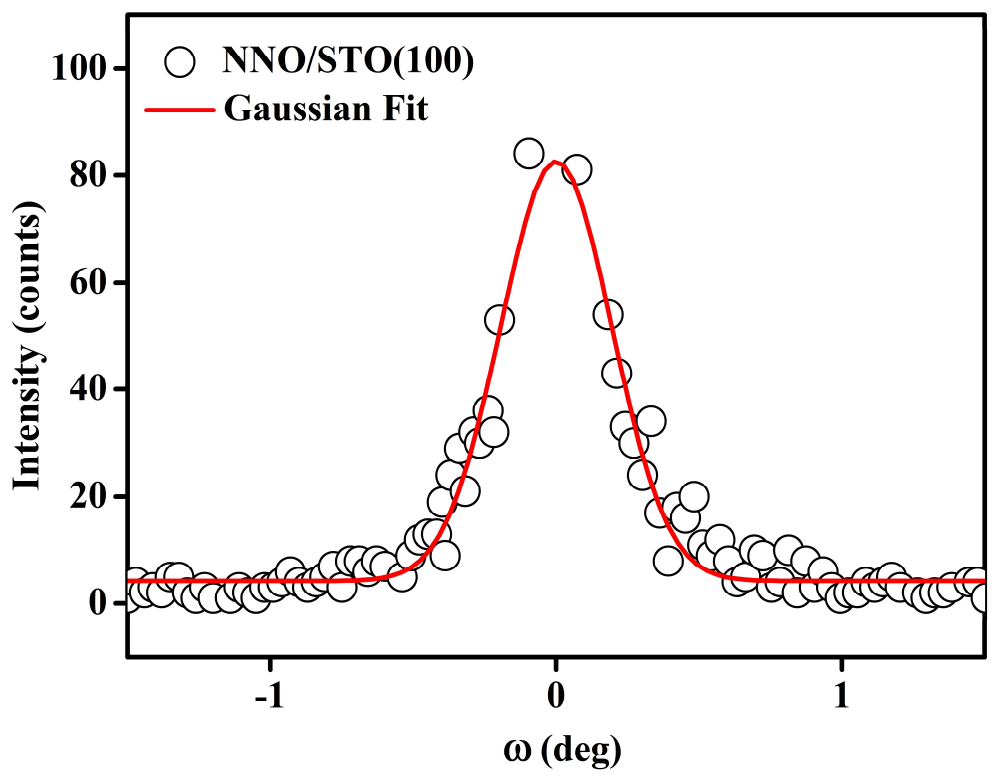

**FIG. S4**: Rocking curves obtained from RSM data for NdNiO3 films grown on SrTiO3(100) substrate and corresponding gaussian fit.

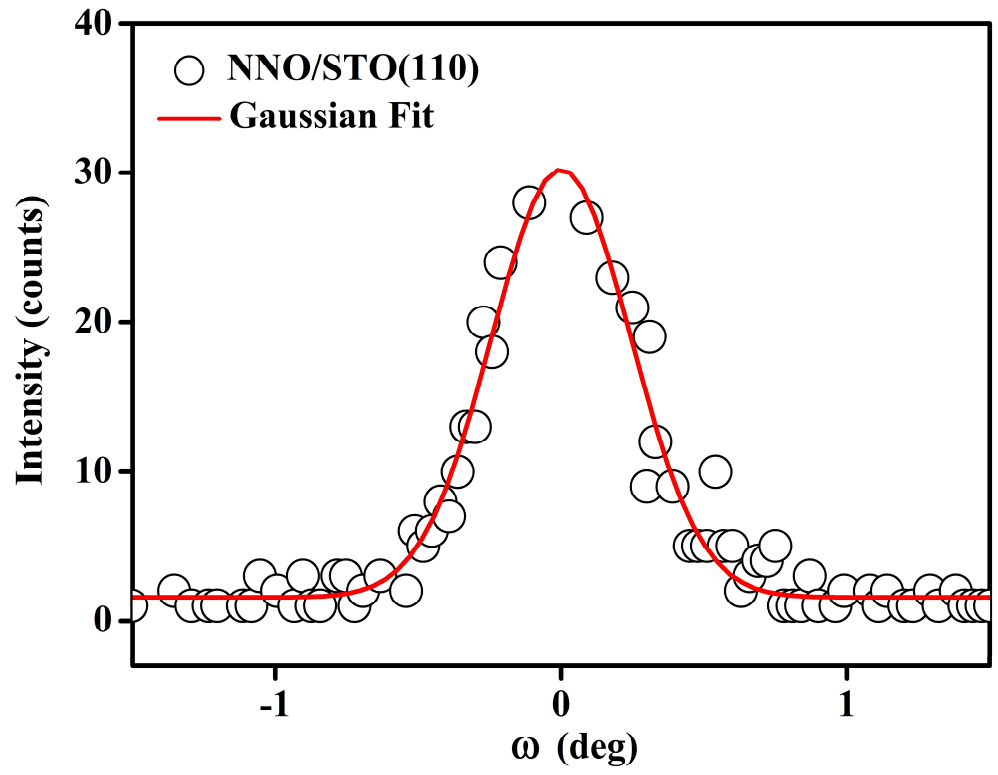

**FIG. S5:** Rocking curves obtained from RSM data for NdNiO3 films grown on SrTiO3(110) substrate and corresponding gaussian fit.

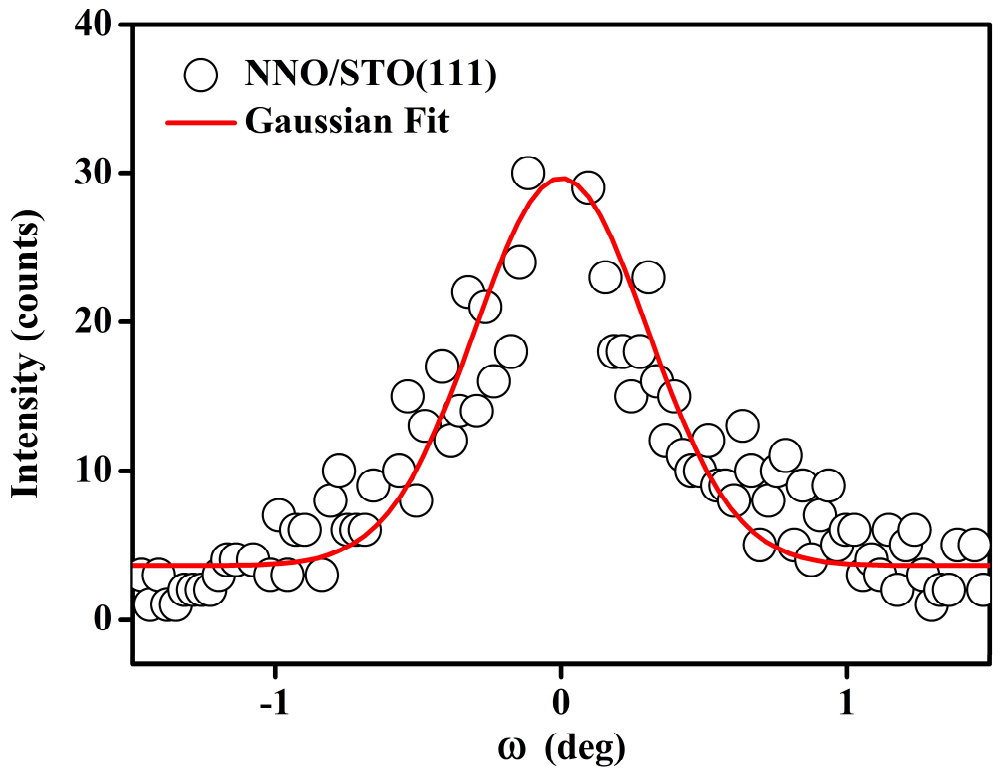

**FIG. S6**: Rocking curves obtained from RSM data for $NdNiO_3$ films grown on $SrTiO_3(111)$ substrate and corresponding gaussian fit

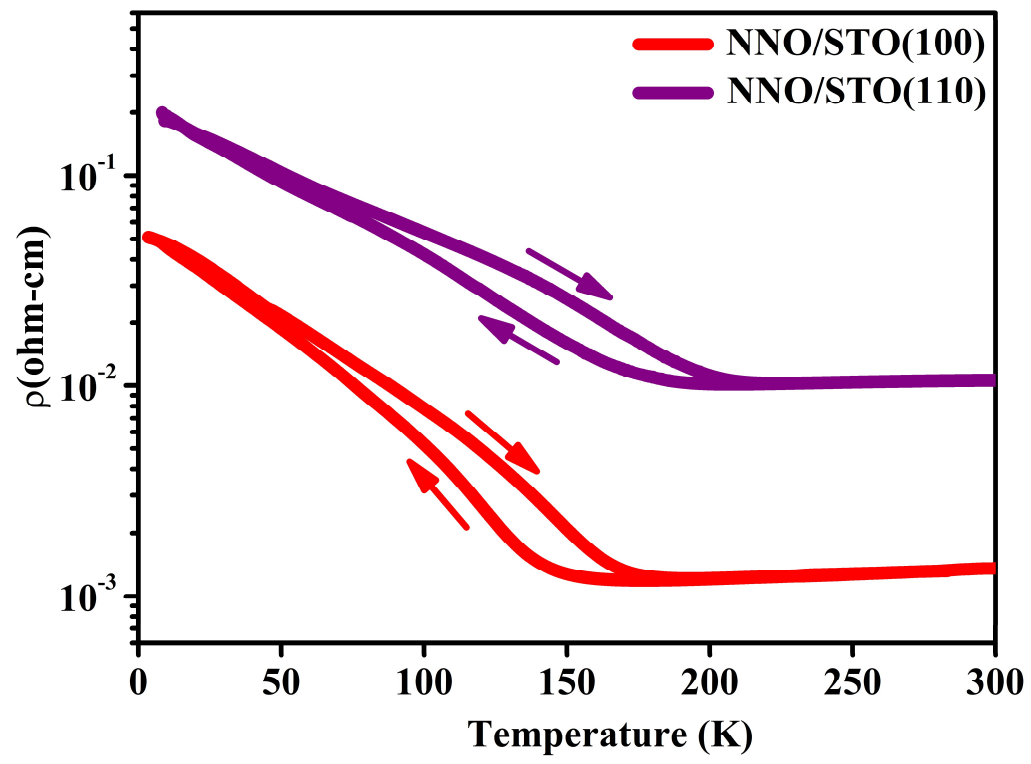

**FIG. S7**: Resistivity ($\rho$) as a function of temperature for $NdNiO_3/SrTiO_3(100)$ and $NdNiO_3/SrTiO_3(110)$ films. The upside and downside arrows represent cooling and heating cycle respectively.

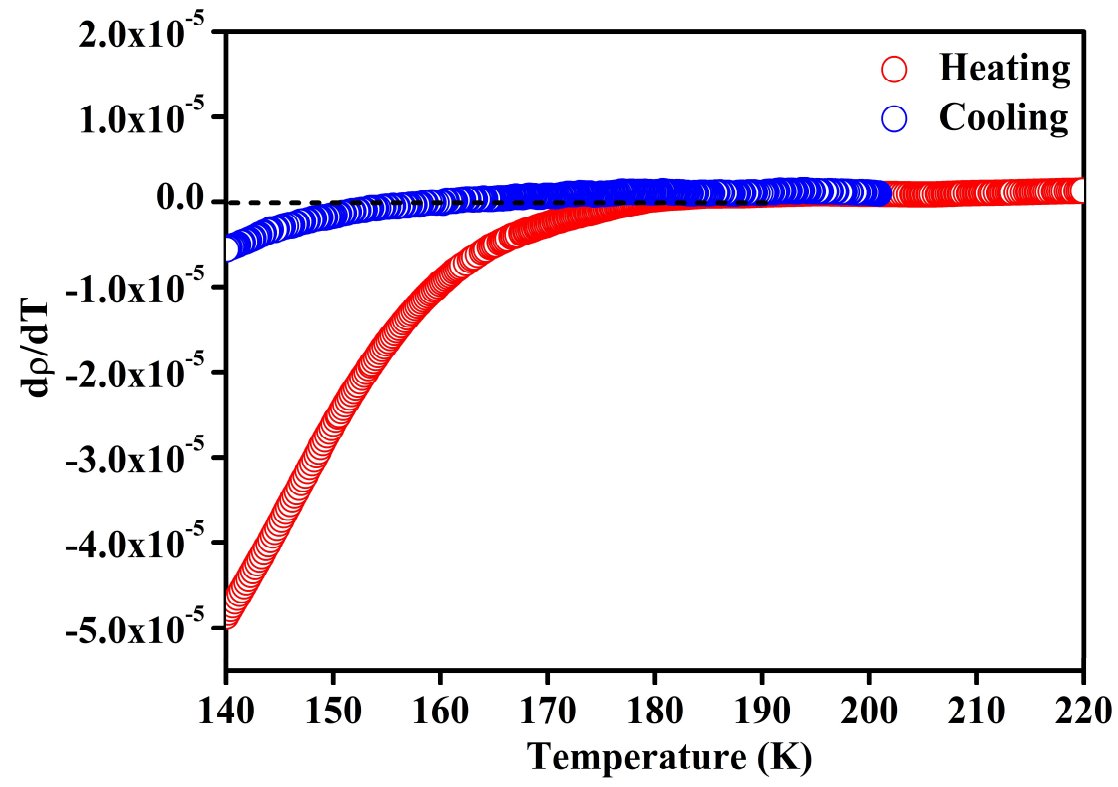

**FIG. S8**: $\frac{d\rho}{dT}$ as a function of T for $NdNiO_3/SrTiO_3(100)$ film. Data shown for heating and cooling cycles.

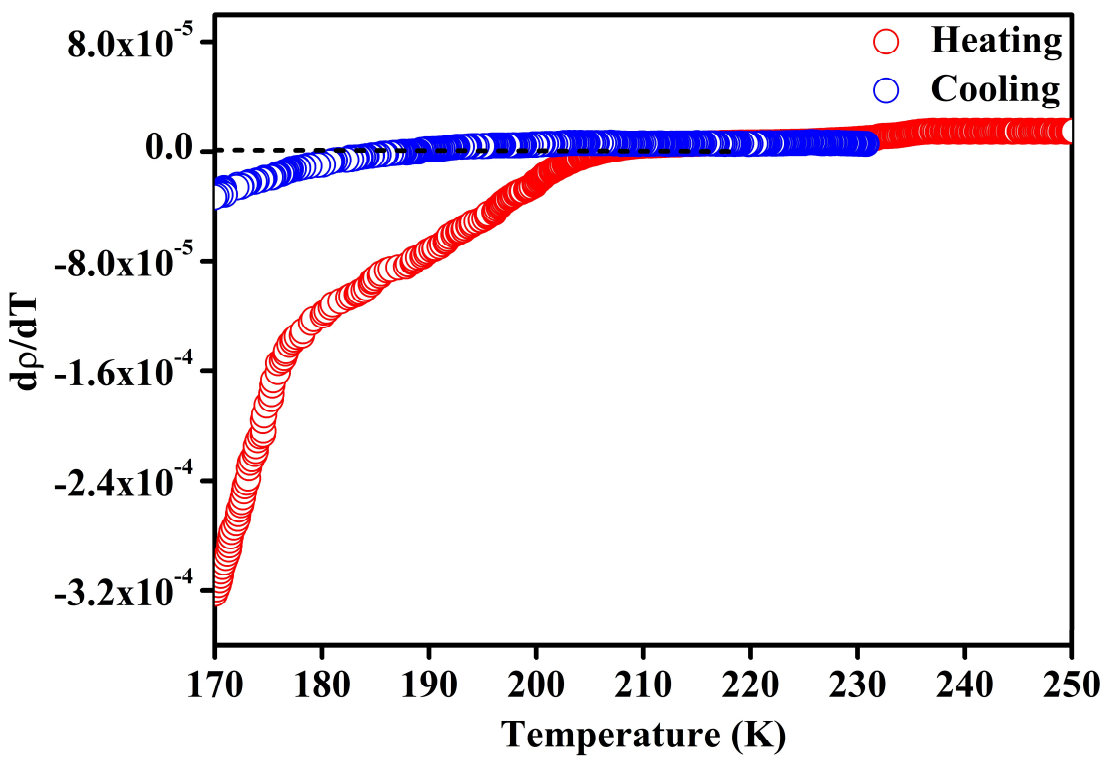

**FIG. S9**: $\frac{d\rho}{dT}$ as a function of T for NdNiO3/SrTiO3(110) film. Data shown for heating and cooling cycles.

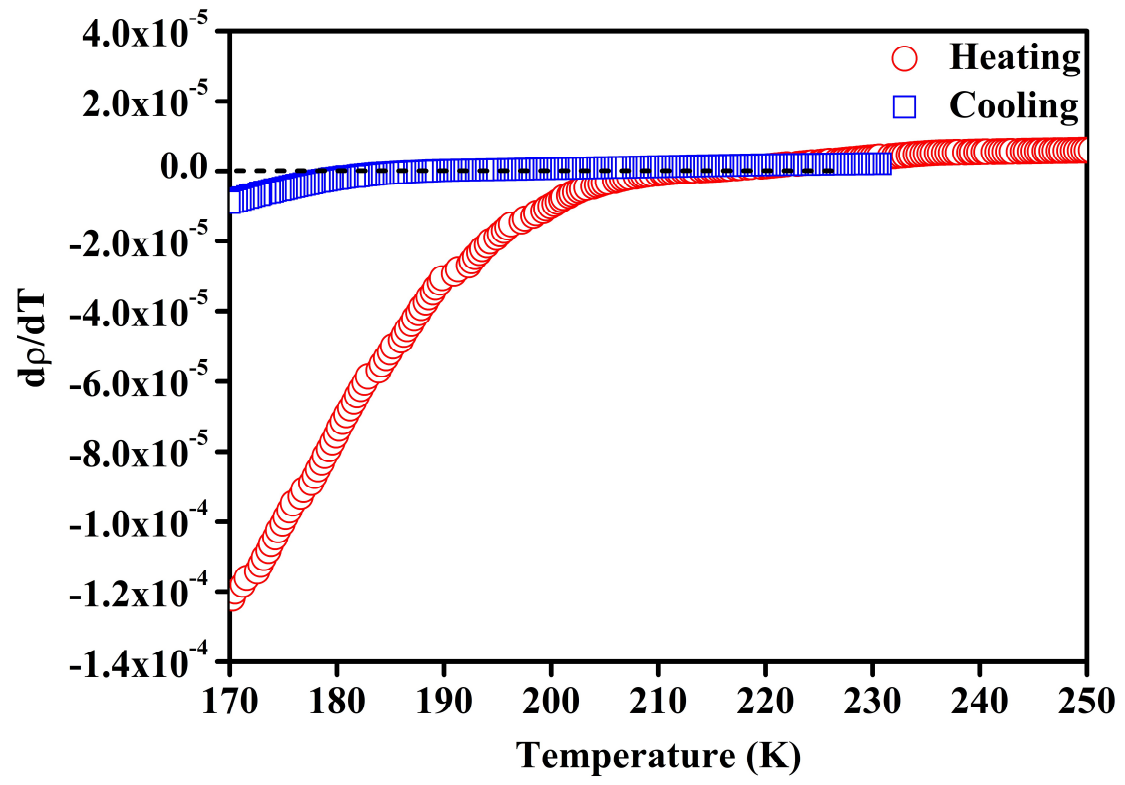

**FIG. S10**: $\frac{d\rho}{dT}$ as a function of T for NdNiO3/SrTiO3(111) film. Data shown for heating and cooling cycles.

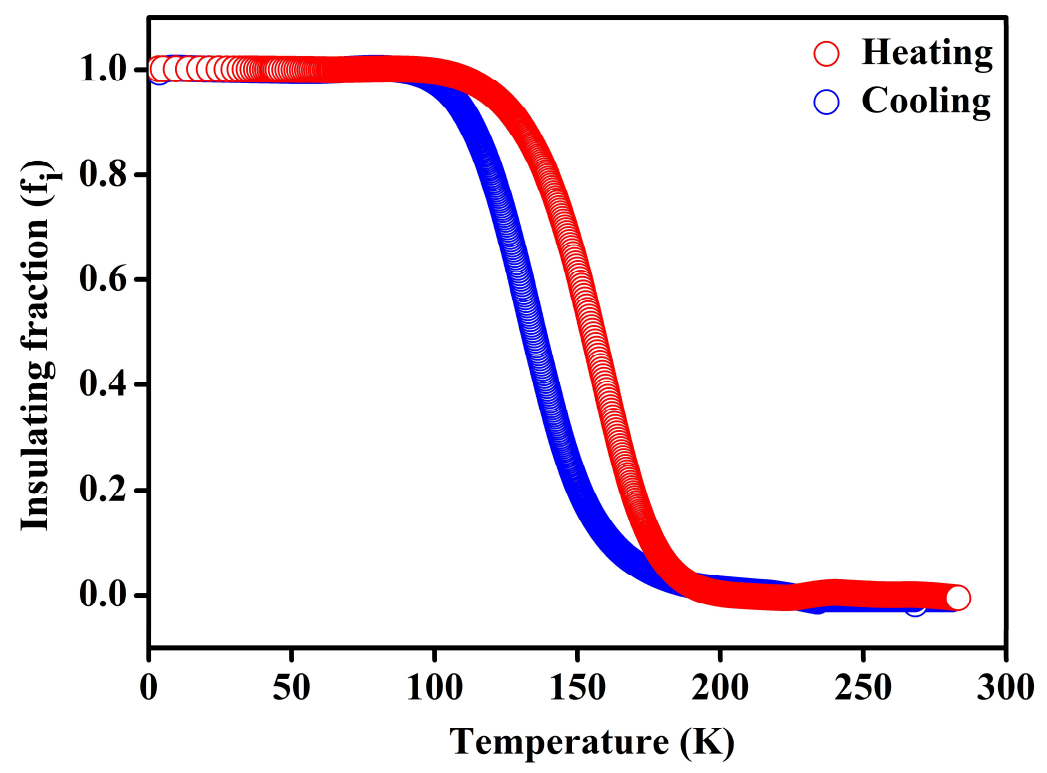

**FIG. S11**: Insulating fraction for NdNiO3 film grown on SrTiO3(100) substrate. Data shown for heating and cooling cycles.

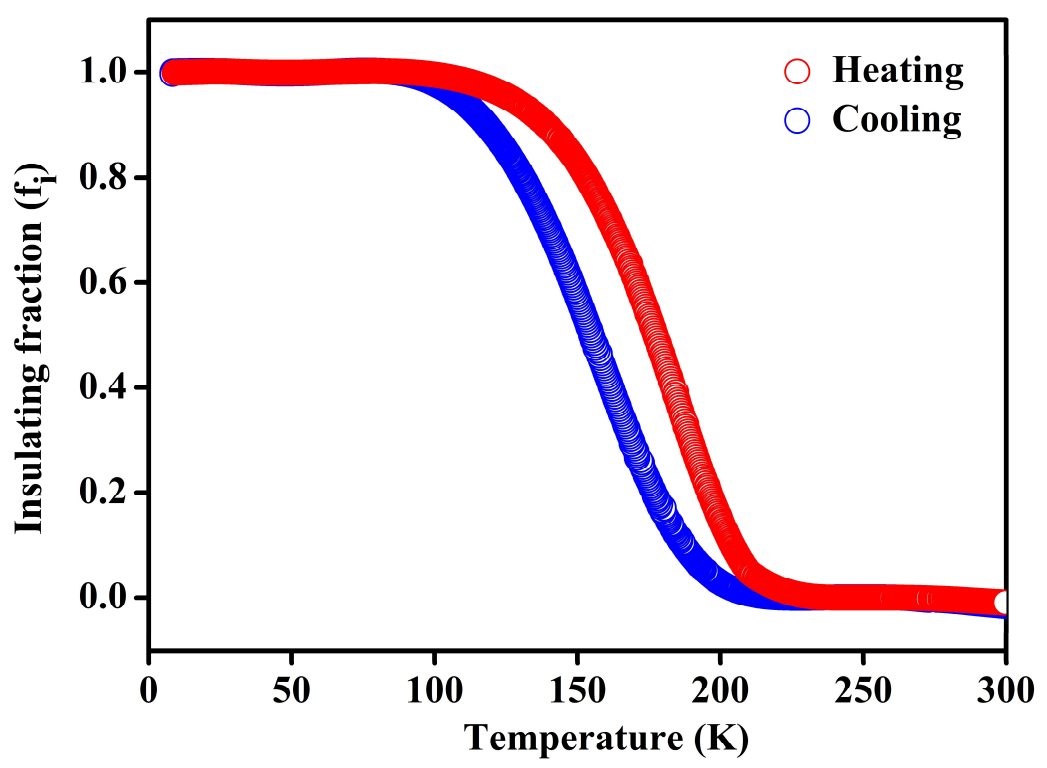

**FIG. S12**: Insulating fraction for NdNiO3 film grown on SrTiO3(110) substrate. Data shown for heating and cooling cycles.

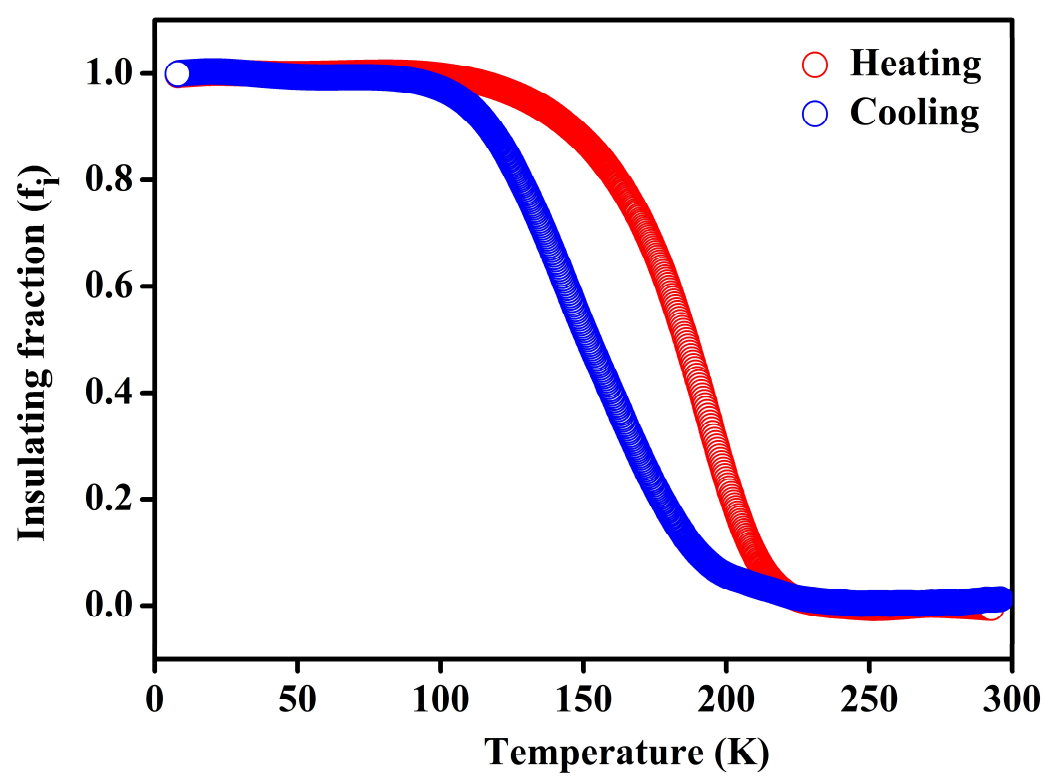

**FIG. S13**: Insulating fraction for NdNiO3 film grown on SrTiO3(111) substrate. Data shown for heating and cooling cycles.

## B. Methods doe measurements of noise spectra

The lock-in amplifier (LIA) based ac modulation noise measurement technique allows us to measure the sample and background noise together. In this technique, a constant alternating current is used to bias the sample [1, 2]. The resulting voltage drop across the sample was then demodulated by the LIA and the output of the LIA (demodulated signal) was fed to a 16 bit analog to digital converter (ADC) card of bandwidth 200 kHz. The voltage fluctuations ($\Delta V(t) = V(t) - < V(t) >$) as a function of time was recorded for nearly 16 minutes (1 Million data points). The obtained time series of the voltage fluctuations were decimated and digitized using digital signal processing techniques. And finally the method of the average periodogram was used to estimate the power spectral density (PSD) [3]. The PSD at the output channel of the lock-in amplifier can be written as [1]:

$$S_v(f, \varphi) = G^2[S_v^{bg}(f - f_0) + I_0^2(f - f_0)S_r(f)\cos^2\varphi] \quad \text{S1}$$

where $S_r(f)$ and $S_v^{bg}(f)$ is the PSD of the sample and background respectively. $\varphi$, $f$, $f_0$, and $G$ represents the phase, measurement frequency, excitation frequency, and gain of the LIA respectively [30]. The dual channel LIA allows us to measure the in-phase and out of phase component altogether and according to equation S1, for $\varphi = 0^0$ (In-phase component), we measure a sum of the sample and the background noise while for $\varphi = 90^0$ (Out of phase component), we measure only background noise. The contribution of the noise from the sample can be extracted after subtracting the background noise from the sum of the total noise.

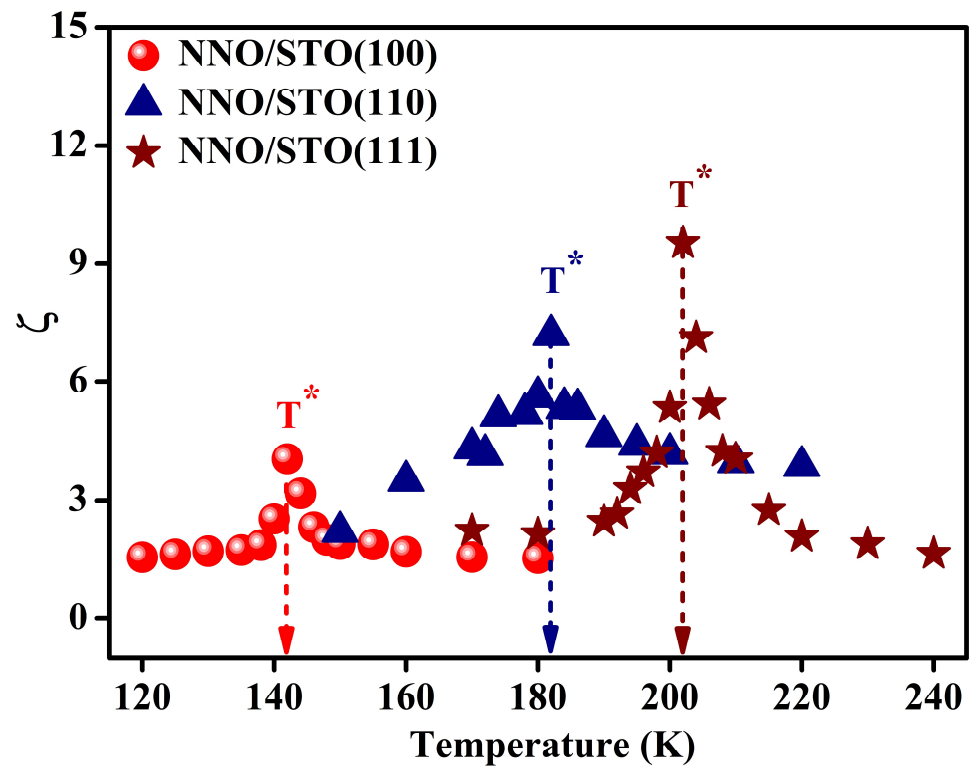

**FIG. S14**: The temperature dependence of the $\zeta$( the ratio of observed white thermal noise spectral power $S_{th}(T)$, and the Johnson-Nyquist noise value of $4k_B TR$ for the resistance value R) for the different samples as a function of temperature for $NdNiO_3$ films grown on $SrTiO_3$ substrates with different orientations. Data are shown for the cooling cycle and the arrow shows the corresponding $T^*$.

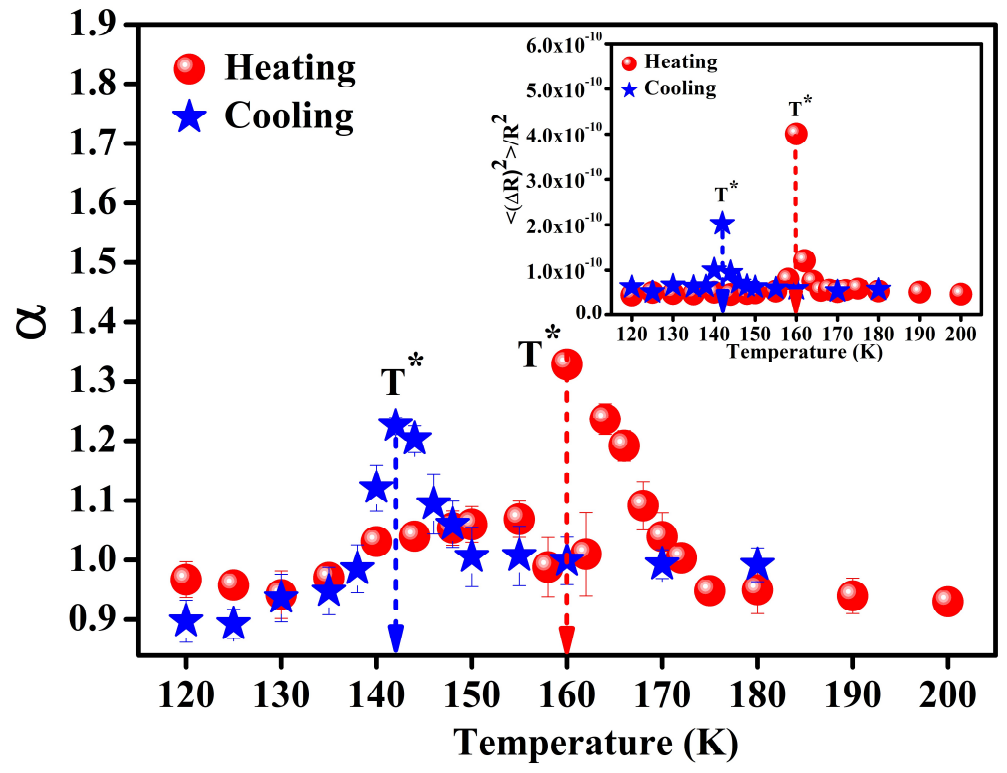

**FIG S15**. Temperature variation of the exponent $\alpha$ for the spectral power density $S_V \sim \frac{1}{f^\alpha}$, for NNO/STO (100) film. Inset: Relative variance $\frac{\langle \Delta R^2 \rangle}{R^2} = \int_{f_{min}}^{f_{max}} df \left(\frac{S_V(f)}{V^2}\right)$. The arrows mark the temperature where $\alpha$ and the variance show peaks.

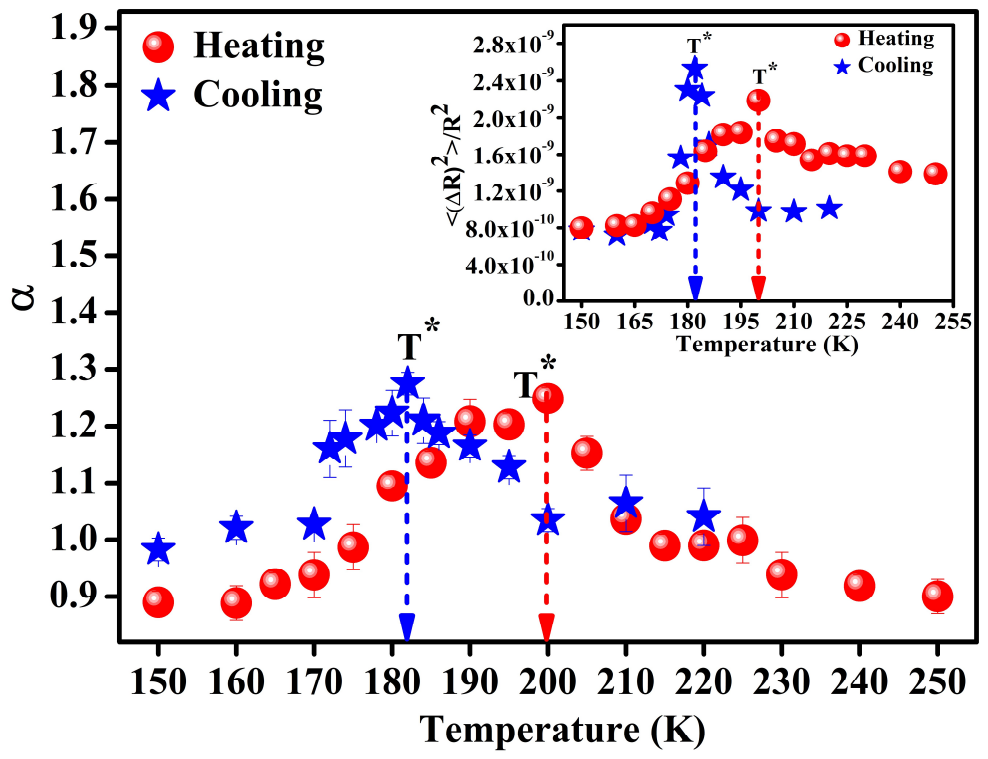

**FIG S16**. Temperature variation of the exponent $\alpha$ for the spectral power density $S_V \sim \frac{1}{f^\alpha}$, for NNO/STO (110) film. Inset: Relative variance $\frac{\langle \Delta R^2 \rangle}{R^2} = \int_{f_{min}}^{f_{max}} df \left(\frac{S_V(f)}{V^2}\right)$. The arrows mark the temperature where $\alpha$ and the variance show peaks.

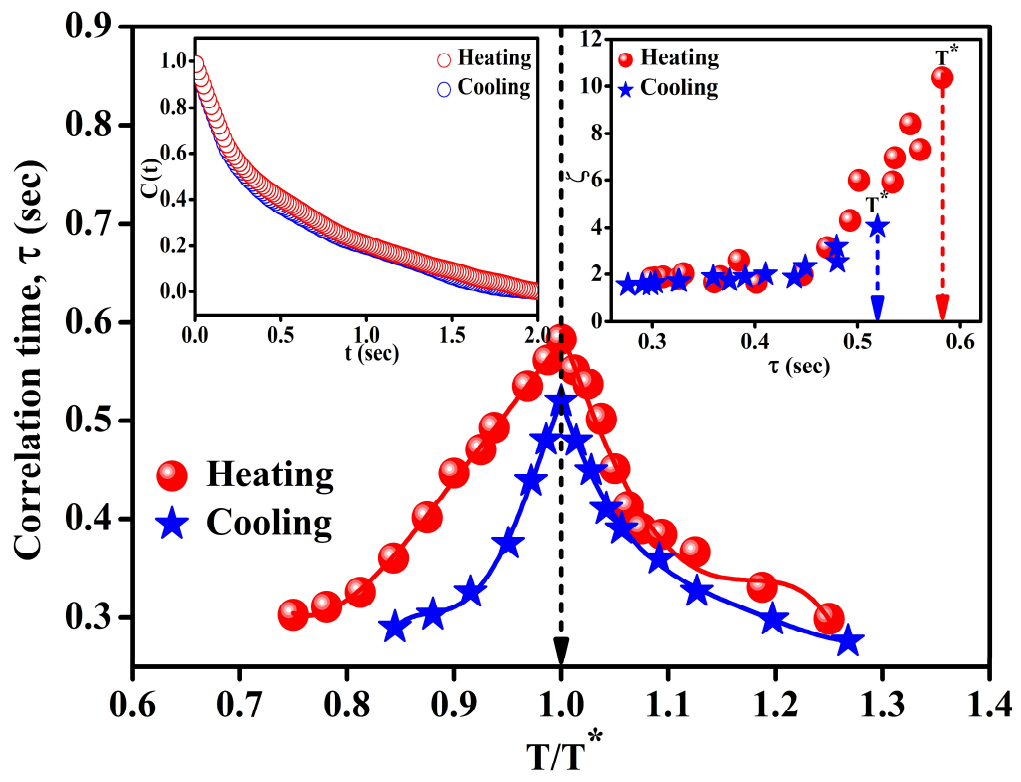

**FIG. S17**: The correlation time $\tau$ as a function of scaled temperature for the NdNiO3/SrTiO3(100) film. The inset at top left corner shows the auto-correlation function as a function of time and inset at top right corner shows a plot of $\zeta$ as a function of $\tau$ . The data are shown for heating and cooling cycle.

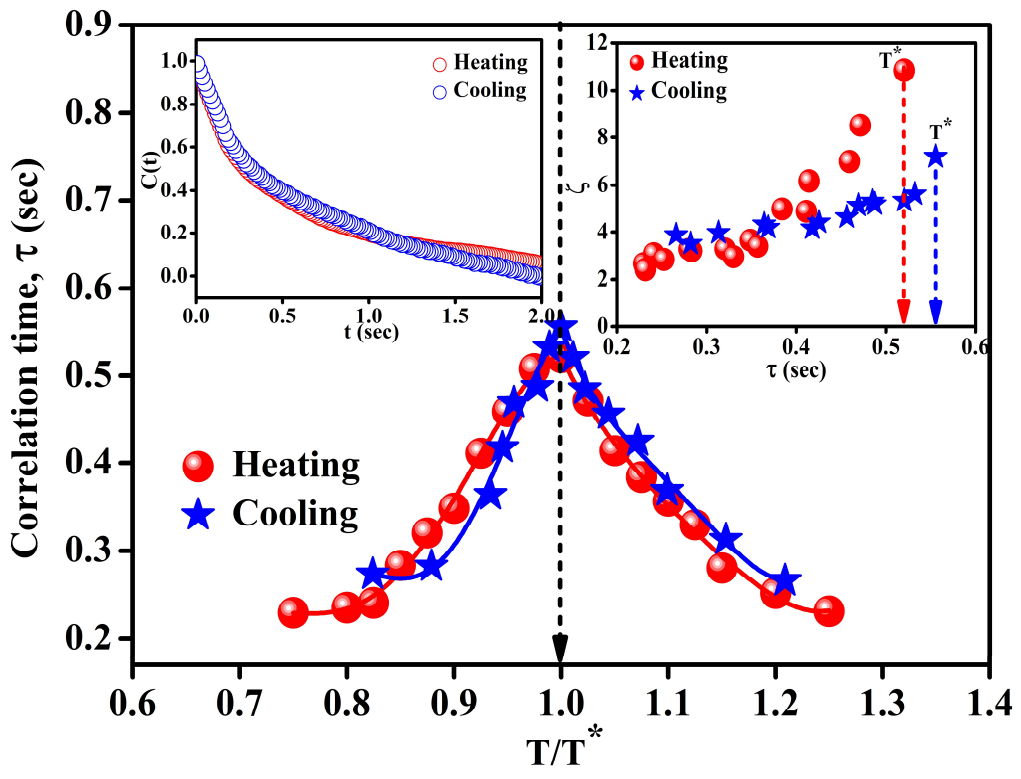

**FIG. S18**: The correlation time $\tau$ as a function of scaled temperature for the NdNiO3/SrTiO3(110) film. The inset at top left corner shows the auto-correlation function as a function of time and inset at top right corner shows a plot of $\zeta$ as a function of $\tau$ . The data are shown for heating and cooling cycle.

## C. Note on 2nd Spectra:

The second spectrum $\Gamma^2$is defined as $= \int_0^{f_H - f_L} S^2(f_2) df_2$ where $S^2(f_2)$ is defined as a normalized second spectrum. The normalization is by the square of the integrated spectral power density (the first spectrum). The definition is given by

$$S^2(f_2) = \frac{\int_0^{\infty} <\Delta V^{(2)}(t)\Delta V^{(2)}(t+\tau)> \cos(2\pi f_2 \tau) d\tau}{[\int_{f_L}^{f_H} S_V(f_1) df_1]^2} \quad \text{S2}$$

Where $f_1$ and $f_2$ are the frequencies associated with first and second spectrum respectively. The spectrum has been calculated within the frequency bandwidth of 0.25 Hz, where $f_l$ is 0.25 Hz, $f_L$ = 0.25 Hz and $f_H$ = 0.5 Hz. For a perfect Gaussian fluctuation, $S^2(f_2) = 3$. The large change in $\Gamma^2$ the second spectrum close to $T_{MI}$ is a signal of correlated non-Gaussian fluctuation.

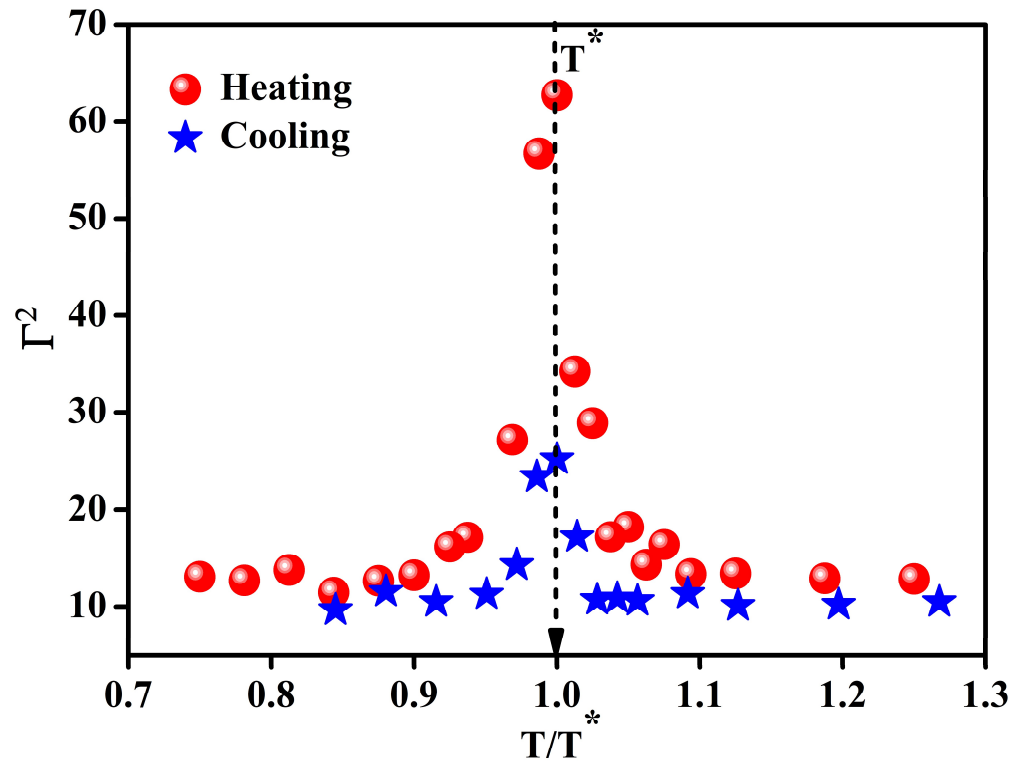

**FIG S19**. The second spectrum $\Gamma^2$ of the noise spectra for NNO/STO (100) film with scaled temperature scale $\frac{T}{T^*}$ for heating and cooling cycle.

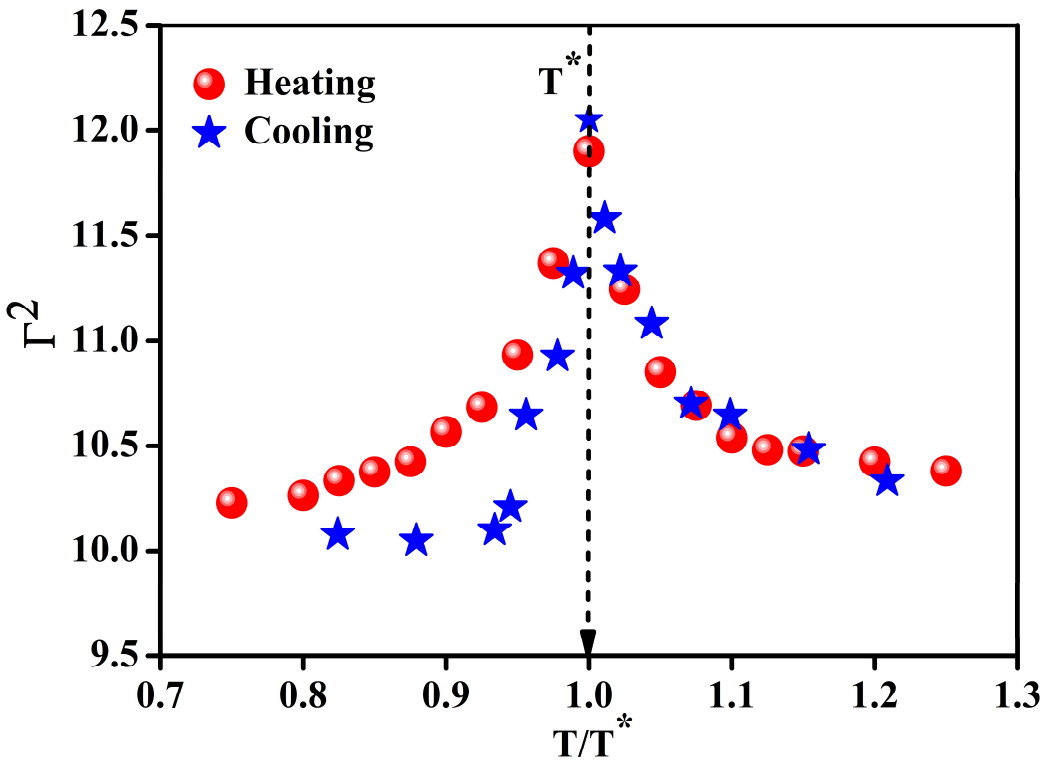

**FIG S20**. The second spectrum $\Gamma^2$ of the noise spectra for NNO/STO (110) film with a scaled temperature scale $\frac{T}{T^*}$ for heating and cooling cycle.

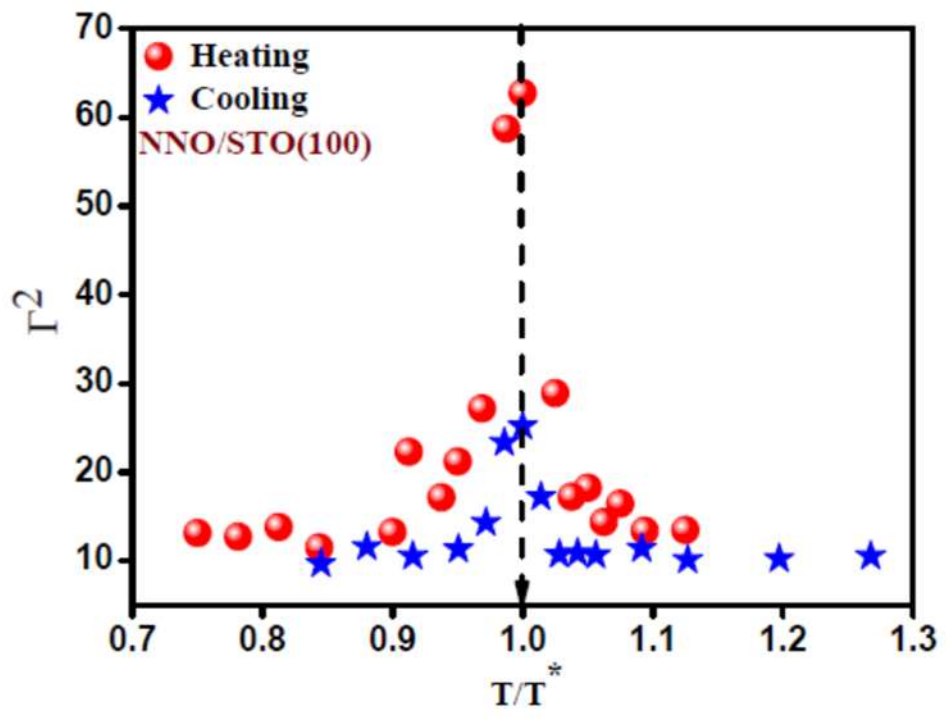

**Figure S21**. The second spectrum $\Gamma^2$ of the noise spectra for NNO/STO (111) film with scaled temperature scale $\frac{T}{T^*}$ for heating and cooling cycle.

## D. Note on Local conductance map (LCMAP)

LCMAP is a spatially resolved image of local tunneling conductance $g(V)$ taken over an aerial range of $0.5\mu m \times 0.5\mu m$ using a UHV Scanning Tunneling Microscope (STM) at a fixed bias V. The data were taken with STM UHV SPM 350 by RHK technology at a base pressure of $10^{-10}$ mbar. The local tunneling conductance $g(\equiv \frac{dI}{dV})$ that provides the information on spatial dependence of the DOS at the Fermi level ($E_F$) has been measured using a modulation method with a small ac bias that is applied to the dc bias $V$ which is used for keeping a fixed height of the tip above the film. The small a.c modulation voltage (<< the dc bias) that has been used to measure the differential tunneling conductance $\frac{dI}{dV}$. Taking a raster scan in the presence of an ac modulation allows one to record the topography as well as spectroscopy data together. The system has a noise level ≤ 10 pA/$\sqrt{Hz}$. The local tunneling conductance map is a contour plot of $g = \frac{dI}{dV}$ taken at a fixed dc bias V=0.5V. The color code gives the corresponding value of $g$(V) with regions of higher tunneling conductance being metallic regions.